\documentclass{PoS}

\usepackage{amsmath,amsfonts,amssymb,epsfig,comment}

\def\twovec[#1,#2]{\left( \begin{array}{c} #1  \\ #2 \end{array} \right)}
\def\twomat[#1,#2][#3,#4]{\left( \begin{array}{cc} #1 & #2 \\ #3 & #4 \end{array} \right)}
\def\threemat[#1,#2,#3][#4,#5,#6][#7,#8,#9]{\left( \begin{array}{ccc} #1 & #2 & #3\\ #4 & #5 & #6 \\ #7 & #8 & #9 \end{array} \right)}

\def\twomat[#1,#2][#3,#4]{\left( \begin{array}{cc} #1 & #2 \\ #3 & #4 \end{array} \right)}
\def\thv[#1,#2,#3]{\left( \begin{array}{c} #1 \\ #2 \\ #3 \end{array} \right)}
\def\twv[#1,#2]{\left( \begin{array}{c} #1 \\ #2 \end{array} \right)}
\def\lagrange{\mathcal{L}}
\def\nn{\nonumber}

\usepackage{color,xspace}
\usepackage{extarrows}

\title{An Emergent Higgs Alignment }

\ShortTitle{An Emergent Higgs Alignment }

\author{\speaker{Karim Benakli}\\
        Sorbonne Universit\'e, UPMC Univ Paris 06, UMR 7589, LPTHE, F-75005, Paris, France \\
CNRS, UMR 7589, LPTHE, F-75005, Paris, France\\
        E-mail: \email{kbenakli@lpthe.jussieu.fr}}

\author{Yifan Chen\\
        Sorbonne Universit\'e, UPMC Univ Paris 06, UMR 7589, LPTHE, F-75005, Paris, France \\
CNRS, UMR 7589, LPTHE, F-75005, Paris, France\\
       E-mail: \email{yifan.chen@lpthe.jussieu.fr}}

\author{{Ga\"etan Lafforgue-Marmet}\\
        Sorbonne Universit\'e, UPMC Univ Paris 06, UMR 7589, LPTHE, F-75005, Paris, France \\
CNRS, UMR 7589, LPTHE, F-75005, Paris, France\\
        E-mail: \email{glm@lpthe.jussieu.fr}}

\abstract{ We explain how an $SU(2)_R$  R-symmetry allows alignment of the vacuum expectation value and the mass eigenstates in the Higgs  bosons squared-mass matrix. More important, we discuss how our parametrization of the breaking of this global symmetry allows to quantitatively apprehend the diverse sources of misalignment, which are kept, by supersymmetry, small.}

\FullConference{Corfu Summer Institute 2018 "School and Workshops on Elementary Particle Physics and Gravity"\\
		(CORFU2018)\\
		31 August - 28 September, 2018\\
		Corfu, Greece}

\begin{document}

Popular extension of the Standard Model (SM) Higgs sector are given by the Two Higgs Doublet Model (2HDM).
The presence of additional  scalars is very much constrained by collider experiments. In fact, the new states could mix with the observable Higgs leading to suppression of its couplings with SM gauge bosons and fermions. These couplings are measured at the LHC, leaving only small room for modifications.  A way to circumvent these modifications is to arrange that   the observed Higgs is \emph{aligned} with the scalar mixing matrix eigenstate acquiring a non-zero vacuum expectation value (v.e.v).  This alignment can be easily achieved by making  the additional scalars heavy enough and decoupling them. But this means that these new states will be kept out of reach at the LHC. A more interesting scenario is to suppress the mixing through somme symmetry that allows then an   \emph{alignment without decoupling}\cite{Gunion:2002zf}. This was realized in  the model of \cite{Antoniadis:2006uj} that was not engineered for as a scenario for supersymmetry breaking, and  the alignment was a "prediction".  This was  discussed  later in \cite{Ellis:2016gxa,Benakli:2018vqz,Benakli:2018vjk}. In \cite{Benakli:2018vqz}, the misalignment was found to remain below the  10$ \%$ level when radiative corrections are taken into account. In \cite{Benakli:2018vjk}, the mechanisms behind this alignment were revealed: they are a combination of a global $SU(2)_R$ symmetry of the quartic potential and diverse cancellations due to supersymmetry. 


The  scalar potential of interest here is the one of \cite{Antoniadis:2006uj} and has been discussed in details  in \cite{Belanger:2009wf}. Alignment in the 2HDM has been discussed by many authors, either as a consequence  of symmetries \cite{Davidson:2005cw,Ivanov:2005hg,Dev:2014yca,Lane:2018ycs} or as a result of tuning the parameters \cite{Bernon:2015qea,Bernon:2015wef,Carena:2013ooa,Carena:2015moc,Haber:2017erd}.  The scalar potential here is different as the model of  \cite{Antoniadis:2006uj} is different from the MSSM or NMSSM for example. Here, the (non-chiral) gauge and Higgs states appear in an $N=2$ supersymmetry sector while the matter states, quarks and leptons, appear in an $N=1$ sector.  Then, it also  differs from $N=2$ extensions of the SM  in \cite{Fayet:1975yi,delAguila:1984qs} where $N=2$ acts on the whole SM states.    One feature of \cite{Antoniadis:2005em,Antoniadis:2006eb,Antoniadis:2006uj,Allanach:2006fy,Belanger:2009wf} to be stressed  is that  gauginos have Dirac masses \cite{Fayet:1978qc,Polchinski:1982an,Hall:1990hq,Fox:2002bu,Benakli:2008pg}.  An $N=2$ extension has implication for Higgs boson physics, as discussed in \cite{Belanger:2009wf,Benakli:2009mk,Amigo:2008rc,Benakli:2010gi,Choi:2010gc,Benakli:2011vb,Benakli:2011kz,Itoyama:2011zi,Benakli:2012cy,Benakli:2014cia,Martin:2015eca,Braathen:2016mmb,Unwin:2012fj,Chakraborty:2018izc,Csaki:2013fla,Benakli:2016ybe}. 

We will review the emergence of the alignment and the origins of the small misalignments.  
 
\section{Higgs alignment from an $SU(2)$ symmetry}

It is trivial to see that an $SU(2)$ symmetry acting on the quartic potential will enforce alignment. In fact, alignment originating due to global symmetries acting on the full Lagrangian have been discussed by a number of authors, for example in \cite{Gunion:2002zf,Dev:2014yca}. A slight difference in our work is that this symmetry acts only on the quartic but not on the quadratic part of the potential. This has for consequence that the ratio of the v.e.v's of the two doublets ($\tan \beta$) is fixed. More interesting is to understand, both qualitatively and quantitatively,  how misalignment is related to the breaking of the global symmetry as this is is the clue for building realistic models. For this purpose, following \cite{Benakli:2018vjk}, we shall parametrize  the measure of misalignment as function of the coefficients of the decomposition under $SU(2)$ of the quartic potential.

It might be more illuminating to the reader to start the discussion  using the by-now standard parametrization of  2HDMs: 
\begin{eqnarray}
V_{EW} &=& V_{2\Phi} +V_{4\Phi} 
\label{decomp2HDM} 
\end{eqnarray}
where
\begin{eqnarray}
V_{2\Phi} &=& m_{11}^2 \Phi_1^\dagger \Phi_1 + m_{22}^2 \Phi_2^\dagger \Phi_2 - [m_{12}^2 \Phi_1^\dagger \Phi_2 + \text{h.c}] \nonumber \\
V_{4\Phi} &=& \frac{1}{2} \lambda_1 (\Phi_1^\dagger \Phi_1)^2 + \frac{1}{2} \lambda_2 (\Phi_2^\dagger \Phi_2)^2 \nonumber \\
& & +  \lambda_3(\Phi_1^\dagger \Phi_1) (\Phi_2^\dagger \Phi_2) + \lambda_4 (\Phi_1^\dagger \Phi_2)(\Phi_2^\dagger \Phi_1) \nonumber \\ 
& & + \left[ \frac{1}{2} \lambda_5 (\Phi_1^\dagger \Phi_2)^2 + [\lambda_6 (\Phi_1^\dagger \Phi_1) + \lambda_7 (\Phi_2^\dagger \Phi_2)] \Phi_1^\dagger \Phi_2 + \text{h.c} \right]\,,  
\label{reparametriz2HDM} 
\end{eqnarray}

We express the parameters $\lambda_i$ as :
\begin{align}
\lambda_i =&\lambda_i^{(0)} + \delta \lambda_i^{(tree)} + \delta \lambda_i^{(rad)}
\label{deltaLs}
\end{align}
where $\lambda_i^{(0)}$  are the leading order tree-level values, $\delta \lambda_i^{(rad)}$ contain the  loop corrections and $\delta \lambda_i^{(tree)}$ the tree-level  threshold corrections , as  first computed in \cite{Belanger:2009wf,Benakli:2012cy}.

We assume that all couplings and vacuum expectation values are real and  CP is conserved and  take:
\begin{align}
\lambda_5^{(0)} =& \lambda_6^{(0)} = \lambda_7^{(0)} =0.
\label{lambdaZeros}
\end{align}
as it is the case below for the particular models of interest for us.

Now, from $\Phi_1$ and $\Phi_2$ we can form  a bi-doublet  $(\Phi_1, \Phi_2)^T$. Here,   $\Phi_1$ and $\Phi_2$ should be seen each as a column with two entries. The weak interaction $SU(2)_w$ acts vertically (on the rows) while another $SU(2)_R$ acts horizontally (on the columns). Here, R stands for R-symmetry as we will discuss below. We will classify the terms in the potential by the way they transform under this $SU(2)_R$ (they are of course invariant under the SM gauge group).

To be invariant under $SU(2)_R$, the quartic potential should be of the form:
\begin{eqnarray}
V_{4\Phi} &=& \, \, \lambda_{|0_1,0>}   |0_1,0\rangle  \quad +  \quad  \lambda_{|0_2,0>}   |0_2,0\rangle
\label{reparam2HDM-SU2R} 
\end{eqnarray}
where we use for $SU(2)_R$ representations the standard notation for spin irreducible representations $|l,m>$. Here:
\begin{eqnarray}
\begin{array}{lll}
|0_1,0\rangle&=& \frac{1}{2}\left[(\Phi^\dagger_1\Phi_1) + (\Phi^\dagger_2 \Phi_2)\right]^2 \, ,
\end{array}
\label{1-1ofSU(2)}
\end{eqnarray}
and 
\begin{eqnarray}
\begin{array}{lll}
|0_2,0\rangle&=& - \frac{1}{\sqrt{12}}\left[\left((\Phi^\dagger_1\Phi_1) - (\Phi^\dagger_2 \Phi_2)\right)^2 \, 
+ 4(\Phi^\dagger_2\Phi_1)(\Phi^\dagger_1\Phi_2)\right]
\end{array}
\label{1-2ofSU(2)}
\end{eqnarray}
and the coefficients are related to the ones in  (\ref{reparametriz2HDM}) by:
\begin{eqnarray}
\lambda_{|0_1,0>} &=&\frac {\lambda_1 + \lambda_2 + 2\lambda_3}{ 4} 
\label{lambdaofSU(2)1}
\end{eqnarray}
and
\begin{eqnarray}
\lambda_{|0_2,0>} = -\frac  {\lambda_1 + \lambda_2 - 2\lambda_3 +4\lambda_4 }{ 4\sqrt{3}} 
\label{lambdaofSU(2)2}
\end{eqnarray}

For the  CP conserving case, the squared-mass matrix in the Higgs basis for the two CP even scalars can be parametrized as  (e.g. \cite{Davidson:2005cw})
\begin{eqnarray}
\mathcal{M}^2_h = \begin{pmatrix}
Z_1 v^2 & Z_6 v^2 \\
Z_6 v^2 & m_A^2 + Z_5 v^2 \end{pmatrix} \, . 
\label{2HDM_mass_matrix}
\end{eqnarray}
where
\begin{align}
Z_1 =& \lambda_1c_\beta^4 + \lambda_2 s_\beta^4 + \frac{1}{2} \lambda_{345} s_{2\beta}^2,   \nn\\
 Z_5 =& \frac{1}{4} s_{2\beta}^2 \left[ \lambda_1 + \lambda_2 - 2\lambda_{345}\right] + \lambda_5  \nn\\
Z_6 =& -\frac{1}{2} s_{2\beta} \left[\lambda_1 c_\beta^2 - \lambda_2 s_\beta^2 - \lambda_{345} c_{2\beta} \right] 
\label{2HDMZ}
\end{align} 
with $\lambda_{345} \equiv \lambda_3 + \lambda_4 + \lambda_5$, and  the pseudo-scalar mass $m_A$ given by
\begin{align}
m_A^2 =& - \frac{m_{12}^2}{s_\beta c_\beta} - \lambda_5 v^2 \qquad \xlongrightarrow{ \lambda_5=0} \qquad  - \frac{m_{12}^2}{s_\beta c_\beta}
\end{align}
Here, we have used again standard definitions:
\begin{eqnarray}
<Re(\Phi_{2}^0)>& =& {v s_\beta }, \qquad <Re(\Phi_{1}^0)>={v c_\beta},  \\
\end{eqnarray}
where:
\begin{eqnarray}
c_\beta &\equiv& \cos \beta,\qquad   s_\beta \equiv \sin \beta, \qquad  t_\beta \equiv \tan\beta \, ,  \qquad   0 \leqslant \beta \leqslant \frac{\pi}{2} \nonumber \\
c_{2\beta} &\equiv& \cos 2\beta,\qquad   s_{2\beta} \equiv \sin 2\beta
\end{eqnarray}

We are particularly interested is the off-diagonal squared-mass matrix coefficient $Z_6$. This measures the amount of misalignment in the absence of decoupling. Following \cite{Benakli:2018vjk}, it can be expressed using the decomposition of the potential in an $SU(2)_R$  basis as
\begin{align}
Z_6 = \frac{1}{2} s_{2\beta} \left[  \sqrt{2} \lambda_{|1,0>} -  \sqrt{6} \lambda_{|2,0>} c_{2\beta}  + (\lambda_{|2,-2>} + \lambda_{|2,+2>}) c_{2\beta}. \right] 
\label{Z6-tree-SU(2)}
\end{align} 
where we used the notation (see \cite{Ivanov:2005hg}):
\begin{eqnarray}
\begin{array}{lll}
|1,0\rangle&=& \frac{1}{\sqrt{2}}\left[(\Phi^\dagger_2\Phi_2) - (\Phi^\dagger_1 \Phi_1)\right]
\left[(\Phi^\dagger_1\Phi_1) + (\Phi^\dagger_2 \Phi_2)\right] \\[1.5mm]
|2,0\rangle&=& \frac{1}{\sqrt{6}}\left[(\Phi^\dagger_1\Phi_1)^2 + (\Phi^\dagger_2 \Phi_2)^2 
 - 2(\Phi^\dagger_1\Phi_1)(\Phi^\dagger_2\Phi_2) -  2(\Phi^\dagger_1\Phi_2)(\Phi^\dagger_2\Phi_1)\right] \\[1.5mm]
|2,+2\rangle&=& (\Phi^\dagger_2\Phi_1)(\Phi^\dagger_2 \Phi_1) \\ [1.5mm]
|2,-2\rangle&=& (\Phi^\dagger_1\Phi_2)(\Phi^\dagger_1 \Phi_2) 
\end{array}
\label{5ofSU(2)}
\end{eqnarray}
The coefficients $\lambda_{|l,m>}$  in (\ref{Z6-tree-SU(2)}) can be expressed as function of the $\lambda_i$ as:
\begin{eqnarray}
\lambda_{|1,0>} &=&  \frac {\lambda_2 - \lambda_1 }{ 2\sqrt{2}}\, \qquad \xlongrightarrow{SU(2)_R}  \qquad  0 \nn \\ [1.5mm]
\lambda_{|2,0>}  &= &\frac {\lambda_1 + \lambda_2 - 2\lambda_3 - 2\lambda_4 }{ \sqrt{24}}\, \qquad \xlongrightarrow{SU(2)_R}  \qquad  0  \nn  \\ [1.5mm]
\lambda_{|2,+2>} &=& \frac{\lambda_5^*}{2}\, \quad \xlongrightarrow{{\rm leading\,  order}}  \quad  0, \qquad 
 \lambda_{|2,-2>} = \frac{\lambda_5}{2}\, \quad \xlongrightarrow{{\rm leading\,  order}}  \quad  0.
\label{lambdaofSU(2)}
\end{eqnarray}
We see that assuming the invariance of the quartic potential under $SU(2)_R$ implies relations between the different $\lambda_i$ which in turn lead to an automatic alignment.  

Note that the expression of $Z_6$ in (\ref{Z6-tree-SU(2)}) shows that the full non-abelian structure is needed and the breaking of $SU(2)_R$ even just to its abelian sub-group spoils the alignment. 
%
The first place where we look for R-symmetry breaking is the quadratic part of the scalar potential. This can be written as:
\begin{eqnarray}
V_{2\Phi} &=& \, \,   \frac{ m_{11}^2  + m_{22}^2}{\sqrt{2}}\times  \frac{1}{\sqrt{2}} \left[  (\Phi_1^\dagger \Phi_1) + (\Phi_2^\dagger \Phi_2)  \right]
 \nonumber \\
& & +  \frac{ m_{11}^2  - m_{22}^2}{\sqrt{2}}\times  \frac{1}{\sqrt{2}} \left[  (\Phi_1^\dagger \Phi_1) - (\Phi_2^\dagger \Phi_2)  \right] \nonumber \\
& & - [m_{12}^2 \Phi_1^\dagger \Phi_2 + \text{h.c}] 
\label{reparam2HDM} 
\end{eqnarray}
where  the first line is the only $SU(2)_R$ invariant part. The potential minimization gives (e.g. \cite{Haber:1993an}):
\begin{eqnarray}
0 &=& m_{11}^2 - t_{\beta} m_{12}^2  + \frac{1}{2} v^2 c_\beta^2 (\lambda_1 +  \lambda_6 t_\beta + \lambda_{345} t_\beta^2 + \lambda_7 t_\beta^2)    \nn  \\
0 &=& m_{22}^2 - \frac{1}{t_{\beta}} m_{12}^2  + \frac{1}{2} v^2 s_\beta^2 (\lambda_2 +  \lambda_7 \frac{1}{t_\beta} + \lambda_{345} \frac{1}{t_\beta^2} + \lambda_6 \frac{1}{t_\beta^2})  
\label{EqtsMotion12} 
\end{eqnarray} 
With $\lambda_1= \lambda_2 = \lambda_{345}\equiv \lambda$ and $\lambda_6= \lambda_7 = 0$, from (\ref{lambdaofSU(2)}), this leads to 

\begin{eqnarray}
0 &=& m_{11}^2 - t_{\beta} m_{12}^2  + \frac{1}{2} \lambda v^2     \\
0 &=& m_{22}^2 - \frac{1}{t_{\beta}} m_{12}^2  + \frac{1}{2} \lambda v^2  
\label{EqtsMotion2} 
\end{eqnarray} 
which, when subtracted one from the other, give (for $s_{2\beta} \neq 0$) 
\begin{eqnarray}
0 &=& \frac{1}{2} (m_{11}^2 - m_{22}^2) s_{2\beta} + m_{12}^2  c_{2\beta} \equiv Z_6 v^2
\label{EqtsMotion3} 
\end{eqnarray} 
Given that the masses $m_{11}$,  $m_{22}$  and $ m_{12}$ are arbitrary, this equation fixes $t_{\beta}$ and implies an \textit{automatic alignment without decoupling}.


\section{A model with naturally realized $SU(2)_R$ symmetry}

There is a simple and elegant way to make the quartic potential $SU(2)_R$ invariant: identify the $SU(2)_R$ as the R-symmetry of $N=2$ supersymmetry and make of the two Higgs doublets the scalar components of a single hypermultiplet $(\Phi_1, \Phi_2)^T$. Considering only these scalars, the $SU(2)_R$  acts now as a Higgs family global symmetry \cite{Davidson:2005cw,Ivanov:2005hg}. There is  a slight complication that one has to deal with though.  This quartic potential receives contributions from  $D$-terms thus we must also extend the $N=2$ supersymmetry to the gauge sector.  This then implies the presence of chiral superfields in the adjoint representations of SM gauge group, we denote a singlet $\mathbf{S}$ and an $SU(2)$ triplet  $\mathbf{T}$. Their scalar components enter in the potential and therefore what we first construct this way is not a 2HDM but rather an extension with a singlet and and a triplet. We proceed then by given these extra fields very heavy masses to decouple them from the light spectrum and finally get to a 2HDM effective potential. We shall describe here some of the main aspects of this model.

We start by defining the new fields:
\begin{eqnarray}
S &=& \frac{S_R + iS_I}{\sqrt{2}} \\
T &=& \frac{1}{2} 
\begin{pmatrix}
T_0 & \sqrt{2} T_+ \\
\sqrt{2}T_- & -T_0 
\end{pmatrix} \,, \qquad T_i = \frac{1}{\sqrt{2}}\, (T_{iR} + i T_{iI})  \quad {\rm  with}  \quad i=0,+, -
\end{eqnarray}
They contribute to the superpotential  by promoting the gauginos to Dirac fermions, but also by generating new Higgs interactions through:

\begin{eqnarray}
W = && \sqrt{2} \, \mathbf{m}_{1D}^\alpha \mathbf{W}_{1\alpha} \mathbf{S} + 2 \sqrt{2} \, \mathbf{m}_{2D}^\alpha \text{tr} \left( \mathbf{W}_{2\alpha} \mathbf{T}\right)  + \frac{M_S}{2} \mathbf{S}^2+ \frac{\kappa}{2} \mathbf{S}^3 + M_T \, \text{tr} (\mathbf{TT}) \,  \nonumber \\ 
 && + \mu \,  \mathbf{H_u} \cdot \mathbf{H_d}+ \lambda_S S \, \mathbf{H_u} \cdot \mathbf{H_d} + 2 \lambda_T \, \mathbf{H_d} \cdot \mathbf{T H_u} \,,
\end{eqnarray} 
\noindent where the Dirac masses are parametrized by spurion superfields $\mathbf{m}_{\alpha i D} = \theta_\alpha m_{iD} $ where $\theta_\alpha$ are the Grassmannian superspace coordinates. The $\lambda_{S,T}$ are not arbitrary as $N=2$ supersymmetry  implies 
\begin{align}
\lambda_S= \frac{1}{\sqrt{2}} g_Y, \qquad \lambda_T =\frac{1}{\sqrt{2}} g_2
\label{LSTN2}
\end{align}
where $g_Y$ and $g_2$ stand for the hyper-charge and $SU(2)$ gauge couplings, respectively. The Higgs potential gets also contributions from soft supersymmetry breaking terms. We chose for simplicity the parameters to be real and we write
\begin{align}
\lagrange_{\rm soft} =& m_{H_u}^2 |H_u|^2 + m_{H_d}^2 |H_d|^2 + B{\mu} (H_u \cdot H_d + \text{h.c})  \nonumber \\ 
 & + m_S^2 |S|^2 + 2 m_T^2 \text{tr} (T^{\dagger} T) + \frac{1}{2} B_S \left(S^2 + h.c\right)+  B_T\left(\text{tr}(T T) + h.c.\right)  \label{soft} \\
& + A_S \left(S H_u \cdot H_d + h.c \right) + 2 A_T  \left( H_d \cdot T H_u + h.c \right) + \frac{A_\kappa}{3} \left( S^3 + h.c. \right) + A_{ST} \left(S \mathrm{tr} (TT) + h.c \right). \nn 
\end{align}

A peculiar 2HDM, with an extended set of light charginos and neutralinos, is obtained by integrating out of the adjoint scalars. The details of this potential were discussed in \cite{Belanger:2009wf}. The result can be mapped to (\ref{reparametriz2HDM}) after the identification
\begin{align}
\Phi_2 = H_u, \qquad \Phi_1^i = -\epsilon_{ij} (H_d^j)^* \Leftrightarrow \twv[H_d^0,H_d^-] = \twv[\Phi_1^0,-(\Phi_1^+)^*] 
\end{align}
from which we can now read
\begin{eqnarray}
m_{11}^2 &=& m_{H_{d}}^2 + \mu^2, \qquad m_{22}^2 = m_{H_{u}}^2 + \mu^2, \qquad m_{12}^2 = B\mu . 
\label{2HDM_params}
\end{eqnarray}
and 
\begin{align}
\lambda_1^{(0)}= \lambda_2^{(0)} =& \frac{1}{4} (g_2^2 + g_Y^2)  \nn \\
\lambda_3^{(0)} =&   \frac{1}{4}(g_2^2 - g_Y^2) + 2 \lambda_T^2  \qquad \xlongrightarrow{N=2}  \qquad  \frac{1}{4}(5 g_2^2 - 
g_Y^2)  \nn  \\
\lambda_4 ^{(0)}=& -\frac{1}{2}g_2^2 + \lambda_S^2 - \lambda_T^2  \qquad \xlongrightarrow{N=2} \qquad - g_2^2 + \frac{1}{2} g_Y^2\nn\\
\lambda_5 =& \, \, \lambda_6 = \lambda_7 =0.
\label{EQ:MDGSSMTree}
\end{align}
as given in \cite{Belanger:2009wf,Benakli:2018vqz}.

Again, restricting to the case of CP conserving Lagrangian, the two CP even scalars have squared-mass matrix (\ref{2HDM_mass_matrix})
with
\begin{align}
Z_1 & \qquad \xlongrightarrow{N=2} \qquad \frac{1}{4} (g_2^2 + g_Y^2)  \nn\\
 Z_5 & \qquad \xlongrightarrow{N=2}  \qquad 0 \nn\\
Z_6 &  \qquad \xlongrightarrow{N=2} \qquad  0.
\label{2HDMZ}
\end{align} 
We use:
\begin{eqnarray}
M_Z^2 & =& \frac{g_Y^2 + g_2^2}{4} v^2 \, ,  \qquad  v \simeq 246 {\rm GeV}  \\
<H_{uR}>& =& {v s_\beta }, \qquad <H_{dR}>={v c_\beta},  \\
 <S_R>&=& v_s \, , \qquad <T_R>=v_t
\end{eqnarray}

Now $m_A$ is given by
\begin{align}
m_A^2 =& - \frac{m_{12}^2}{s_\beta c_\beta} - \lambda_5 v^2 \qquad \xlongrightarrow{N=2} \qquad  - \frac{m_{12}^2}{s_\beta c_\beta}
\end{align}
and squared-mass matrix has eigenvalues:
\begin{align}
m_{h}^2 &= \frac{1}{4} (g_2^2 + g_Y^2) v^2= M_Z^2 \nn\\
 m_{H}^2 &= m_A^2 
\label{ZH}
\end{align} 
while the charged Higgs has a mass
\begin{align}
m_{H^+}^2 =& \frac{1}{2} ( \lambda_5 - \lambda_4) v^2 + m_{A}^2 \qquad \xlongrightarrow{N=2} \qquad    \frac{1}{2}  (g_2^2 - \frac{1}{2} g_Y^2) v^2 + m_{A}^2=3M_W^2-M_Z^2+m_A^2.
\end{align}

Also, the leading-order squared-masses for the real part of the adjoint fields are \cite{Benakli:2011kz}:
\begin{align}
m_{SR}^2 =& m_S^2 + 4 m_{DY}^2 + B_S , \qquad m_{TR}^2 = m_T^2  + 4 m_{D2}^2 + B_T \, .
\label{STscalarmasses}
\end{align}
where we have taken $M_S=M_T=0$.

Let us turn now to the quadratic part of the potential. It can be written as (\ref{reparam2HDM}). Imposing a Higgs family symmetry would have required that both coefficients of the two $SU(2)_R$ non-singlets operators to vanish, therefore $m_{11}^2  = m_{22}^2$ and $m_{12}=0$. First, this would imply $m_{A}^2=0$ which is not a viable feature. Second, 
 the mass parameters in the quadratic potential under $SU(2)_R$ are controlled by the supersymmetry breaking mechanism and this is not expected to preserve the $R$-symmetry.  It was shown in \cite{Benakli:2008pg} that absence of tachyonic directions in the adjoint fields scalar potential implies that in a gauge mediation scenario that either breaking or messenger sectors should not be $N=2$ invariant. Thus, the quadratic potential can not be invariant under $SU(2)_R$.

\section{Misalignment from $R$-symmetry breaking }


We have reviewed how the Higgs alignment is enforced by requiring the invariance of the quartic scalar potential under $SU(2)_R$ symmetry. Moreover, we have exhibited a model when this $SU(2)_R$ symmetry is naturally present. However, at the electroweak scale this symmetry is not realized and therefore it is important to investigate the exact correlation between misalignment and the $SU(2)_R$ breaking.

We have seen that the quartic scalar potential can be recasted as:
\begin{eqnarray}
V_{4\Phi} &=& \, \, \sum_{j,m} \lambda_{|j,m>} \times   |j,m\rangle
\label{reparam2HDM-2} 
\end{eqnarray}
where $|j,m\rangle$ are the irreducible representations of $SU(2)_R$.  Also, with $\lambda_5 =0$, tthe misalignment is measured by
\begin{align}
Z_6 = \frac{1}{2} s_{2\beta} \left[  \sqrt{2} \lambda_{|1,0>} - \sqrt{6} \lambda_{|2,0>} c_{2\beta}  \right] 
\label{Z6-tree-SU(2)-2}
\end{align}

The $\lambda_{|i,0>}$ are generated or corrected by higher order contributions  to the scalar potential. We start with two important remarks.

First, the scalars $S$ and $T$ are singlets of $SU(2)_R$ and have  interactions with the two Higgs doublets preserve $SU(2)_R$. As a consequence,  their loop integration will not lead to $SU(2)_R$  breaking, no contribution to $Z_6$ and no misalignment at leading order.  This was obtained by explicit calculations of the loop diagrams in Eq. (3.5) of \cite{Benakli:2018vqz} where summing up different contributions to $Z_6$, they cancel out. This result is now easily understood as a consequence of the $SU(2)_R$ symmetry.

Second, let's call  $D^a$ for the gauge fields $A^a$ and $F_\Sigma^a$ the auxiliary fields for the adjoint scalars $\Sigma^a \in \{ S,T^a \}$ of $U(1)_Y$ and $SU(2)$ respectively. Then :
\begin{align}
( \quad Re(F_\Sigma^a) \quad ,  \qquad D^a \quad , \qquad  {Im(F_\Sigma^a)} \quad )
\label{Auxil}
\end{align}
form a triplet of $SU(2)_R$. This is at the origin of the relations $\lambda_S=  g_Y/{\sqrt{2}}$ and $ \lambda_T = g_2/{\sqrt{2}}$ in eq. (\ref{LSTN2}). The violation of these relations by quantum effects translates into breaking of $SU(2)_R$. Loops of the adjoint scalar fields $S$ and $T^a$ do not lead to any contribution at leading order as the couplings $\lambda_S$ and $\lambda_T$ are still given by their $N=2$ values. However other loop corrections can give sizable contributions. For instance, the chiral matter  lead to radiative corrections that take $\lambda_S$ and $ \lambda_T $  away from their $N=2$ values. As $\lambda_1$ and $\lambda_2$ are affected in the same way, we find $\delta \lambda_{|1,0>}^{(2\rightarrow 1)} =0$, and using (\ref{lambdaofSU(2)}),  this $N=2 \rightarrow N=1$ breaking leads to:

 \begin{eqnarray}
\delta  Z_6^{(2\rightarrow 1)} &=& \frac{\sqrt{6}}{2}  \, \,  s_{2\beta}  \, \,  c_{2\beta}   \, \,  \delta   \lambda_{|2,0>}^{(2\rightarrow 1)} \nn \\
&= &  -  \frac{1}{2}  \, \,\frac{ t_\beta (t_\beta^2-1)}{(1+ t_\beta^2)^2}  \, \,\left[ (2 \lambda_S^2- g_Y^2  )  + ( 2 \lambda_T^2 -g_2^2 ) \right] 
\label{Z6N2-1}
\end{eqnarray}

The  difference in Yukawa couplings to the two Higgs doublets as well as the $N=1 \rightarrow N=0$ supersymmetry breaking, both lead to further breaking of the $SU(2)_R$ symmetry. For $t_\beta \sim \mathcal{O}(1)$, this misalignment is dominated by the contributions to $\lambda_2$ from stop loops, due to the large Yukawa coupling. It is computed to be: 
\begin{align}
 \delta  \lambda_2 \sim &  \frac{3y_t^4}{8\pi^2} \log \frac{m_{\tilde{t}}^2}{Q^2} 
\label{toplambda2}
\end{align}
Here $Q$, $y_t$, $m_{\tilde{t}}$ are the renormalisation scale, the top Yukawa coupling and the stop mass, respectively. 
All these contributions sum up to give:
\begin{align}
Z_6  \approx&\frac{0.12}{t_\beta} - \, \,\frac{ t_\beta ( t_\beta^2-1)}{(1+ t_\beta^2)^2}  \left[  ( 2 \lambda_S^2 - g_Y^2)+ ( 2 \lambda_T^2 - g_2^2) \right].
\end{align}

Integrating out the heavy adjoint scalars $S$ and $T$ at tree-level, keeping the Higgs $\mu$-term and the Dirac masses $m_{1D}, m_{2D}$  small , in the sub-TeV region, we get:
\begin{align}
\delta \lambda_1^{(tree)} \simeq & -\frac{\left(g_Y m_{1D} - \sqrt{2} \lambda_S \mu\right)^2}{m_{SR}^2} - \frac{\left(g_2 m_{2D} + \sqrt{2} \lambda_T  \mu\right)^2}{m_{TR}^2}\nn \\
\delta  \lambda_2 ^{(tree)} \simeq &   -\frac{\left(g_Y m_{1D} + \sqrt{2} \lambda_S \mu\right)^2}{m_{SR}^2} - \frac{\left(g_2 m_{2D} - \sqrt{2} \lambda_T  \mu\right)^2}{m_{TR}^2} \nn\\
\delta \lambda_3^{(tree)}  \simeq &  \, \, \, \,     \frac{g_Y^2 m_{1D}^2 - 2\lambda_S^2 \mu^2}{m_{SR}^2}-  \frac{g_2^2 m_{2D}^2 - 2\lambda_T^2 \mu^2}{m_{TR}^2} \nn  \\
\delta \lambda_4 ^{(tree)} \simeq & \, \, \, \,  \frac{2 g_2^2 m_{2D}^2 - 4 \lambda_T^2 \mu^2}{m_{TR}^2}\,, \nn\\
\label{deltaLambdaTree}
\end{align}
These means that the quartic potential has corrections of the form:
 \begin{eqnarray}
\delta V_{4\Phi}^{(tree)}  &=& \delta \lambda_{|0_1,0>}^{(tree)}  |0_1,0\rangle  +   \delta \lambda_{|0_2,0>}^{(tree)}   |0_2,0\rangle  +  \delta \lambda_{|1,0>}^{(tree)}  |1,0\rangle  +   \delta \lambda_{|2,0>}^{(tree)}   |2,0\rangle \, .
\label{deltareparam2HDM} 
\end{eqnarray}
and a misalignment arises from the appearance of:
\begin{align}
\delta \lambda_{|1,0>}^{(tree)} \simeq & 2 g_2 \lambda_T  \frac{m_{2D} \mu}{m_{TR}^2}  -2 g_Y \lambda_S  \frac{m_{1D} \mu}{m_{SR}^2}  \nn \\
\simeq & \sqrt{2} g_2^2  \frac{m_{2D} \mu}{m_{TR}^2}  -\sqrt{2} g_Y^2   \frac{m_{1D} \mu}{m_{SR}^2}  \nn \\
\delta  \lambda_{|2,0> }^{(tree)}\simeq &  \sqrt{ \frac{2}{3}} \left[g_Y^2 \frac{ m_{1D}^2 }{m_{SR}^2} + g_2^2 \frac{ m_{2D}^2 }{m_{TR}^2} \right]
\label{deltaLambdas}
\end{align}
These preserve the subgroup $U(1)_R^{(diag)}$. This is because the scalar potential results from  integrating out the adjoints which have zero $U(1)_R^{(diag)}$ charge. For a numerical estimate, we take  $m_{SR} \simeq m_{TR} \simeq 5$  TeV,   $m_{1D}\simeq m_{1D} \simeq \mu \simeq 500$ GeV,  $g_Y\simeq 0.37$ and $g_2 \simeq 0.64$.  This gives 
\begin{align}
\delta \lambda_{|1,0>}^{(tree)} \simeq & 4 \times 10^{-3},  \qquad
\delta  \lambda_{|2,0> }^{(tree)}\simeq 4.5 \times 10^{-3}&  
\label{deltaLambdaTreeSU2R2}
\end{align}
This shows that this contribution to $Z_6$ are numerically negligible.

Even, if this tree-level misalignment is quantitatively small, we would like discuss it here a bit further. We have stated that $S$ and $T$ are singlets of $SU(2)_R$. We have first computed, and understood, how integrating them out in loops does not lead to $SU(2)_R$ breaking. But, then we computed how integration of the same fields at tree-level breaks the $SU(2)_R$ symmetry and leads to misalignment. Why?

In fact, the reason lies in the fact that $(  Re(F_\Sigma^a),   \frac{D^a}{\sqrt{2}} ,  {Im(F_\Sigma^a)} )$ is a triplet of $SU(2)_R$. Let's take for example the case of $S$. The Dirac gauging Lagrangian reads:
\begin{align}
\int d^2 \theta \bigg[ \frac{1}{4} W_{Y\,\alpha}W^\alpha_{Y}+  \sqrt{2} \theta^\alpha  {  m_{D1} \mathbf{S} W_{Y\,\alpha}} + cc  \bigg]+ \int d^4 \theta \,   \sum_j \Phi^\dagger_j e^{Y_jg' V} \Phi_j+ ...  
\end{align}
which leads to
\begin{align}
D_1 = D_Y^{(0)} { \, \,  \longrightarrow} \, \,   D_1  =   - 2 m_{1D} \, \,  {(S + S^*)}\, \,  +\, \, D_Y^{(0)} \qquad \textrm{with}  \qquad D_Y^{(0)}=  - g'\sum_{j} Y_j \varphi_j^* \varphi_j \, .
\end{align}

This shows that a v.e.v of $S$ (or $T$ ) will lead to a  v.e.v. of $D^a$, a component of a triplet of $SU(2)_R$ that breaks spontaneously the symmetry. There is of course no massless Goldstone boson because this global symmetry is already broken explicitly in other parts of the Lagrangian.

%

\section{Conclusions}
%

A tree-level Higgs alignment without decoupling is very easy to achieve. It just goes back to solving simple equations. Beyond tuning different coefficients, we have explained that simply imposing an $SU(2)_R$ symmetry in the quartic potential is enough. The fact that we do not impose the symmetry on the quadratic part of the potential implies than $\tan \beta$, the ratio of the two doublets Higgs vevs is fixed, in contrast for example with the models in \cite{Dev:2014yca}.

Much more interesting, is that we have found a model where not only the required global $SU(2)_R$ symmetry is built in naturally, a consequence of extending supersymmetry from $N=1$ to $N=2$. Writing the off-diagonal elements of the Higgs mass-squared matrix as a linear combination of the coefficients of the decomposition of the potential in spin irreducible representations, we are able to  understand the origin and predict the size of each contribution to misalignment. We are able to show therefore that all the higher order corrections are small and under control. The model has a rich phenomenology that can be tested at LHC as it allows the existence of new light scalars and fermions with electroweak interactions.

\vskip.1in
\noindent
{\bf Acknowledgments}

\noindent {We acknowledge the support of  the Agence Nationale de Recherche under grant ANR-15-CE31-0002 ``HiggsAutomator''. This work is also supported by the Labex ``Institut Lagrange de Paris'' (ANR-11-IDEX-0004-02,  ANR-10-LABX-63).}



\begin{thebibliography} {99}

\bibitem{Gunion:2002zf}
J.~F. Gunion and H.~E. Haber, {\em {The CP conserving two Higgs doublet model:
  The Approach to the decoupling limit}}.
  \href{http://dx.doi.org/10.1103/PhysRevD.67.075019}{Phys. Rev. {\bf D67}
  (2003)  075019},
\href{http://arxiv.org/abs/hep-ph/0207010}{{\tt arXiv:hep-ph/0207010
  [hep-ph]}}.

\bibitem{Antoniadis:2006uj}
I.~Antoniadis, K.~Benakli, A.~Delgado, and M.~Quiros, {\em {A New gauge
  mediation theory}}. Adv. Stud. Theor. Phys. {\bf 2} (2008)  645--672,
\href{http://arxiv.org/abs/hep-ph/0610265}{{\tt arXiv:hep-ph/0610265
  [hep-ph]}}.

\bibitem{Ellis:2016gxa}
J.~Ellis, J.~Quevillon, and V.~Sanz, {\em {Doubling Up on Supersymmetry in the
  Higgs Sector}}. \href{http://dx.doi.org/10.1007/JHEP10(2016)086}{JHEP {\bf
  10} (2016)  086},
\href{http://arxiv.org/abs/1607.05541}{{\tt arXiv:1607.05541 [hep-ph]}}.

\bibitem{Benakli:2018vqz}
K.~Benakli, M.~D. Goodsell, and S.~L. Williamson, {\em {Higgs alignment from
  extended supersymmetry}}.
  \href{http://dx.doi.org/10.1140/epjc/s10052-018-6125-1}{Eur. Phys. J. {\bf
  C78} (2018) no.~8, 658},
\href{http://arxiv.org/abs/1801.08849}{{\tt arXiv:1801.08849 [hep-ph]}}.

\bibitem{Benakli:2018vjk}
  K.~Benakli, Y.~Chen and G.~Lafforgue-Marmet,
  Eur.\ Phys.\ J.\ C {\bf 79} (2019) no.2,  172
  doi:10.1140/epjc/s10052-019-6676-9
  [arXiv:1811.08435 [hep-ph]].
  
\bibitem{Belanger:2009wf}
G.~Belanger, K.~Benakli, M.~Goodsell, C.~Moura, and A.~Pukhov, {\em {Dark
  Matter with Dirac and Majorana Gaugino Masses}}.
  \href{http://dx.doi.org/10.1088/1475-7516/2009/08/027}{JCAP {\bf 0908} (2009)
   027},
\href{http://arxiv.org/abs/0905.1043}{{\tt arXiv:0905.1043 [hep-ph]}}.

%
%

\bibitem{Davidson:2005cw}
S.~Davidson and H.~E. Haber, {\em {Basis-independent methods for the
  two-Higgs-doublet model}}.
  \href{http://dx.doi.org/10.1103/PhysRevD.72.099902,
  10.1103/PhysRevD.72.035004}{Phys. Rev. {\bf D72} (2005)  035004},
  \href{http://arxiv.org/abs/hep-ph/0504050}{{\tt arXiv:hep-ph/0504050
  [hep-ph]}}.
[Erratum: Phys. Rev.D72,099902(2005)].

\bibitem{Ivanov:2005hg}
I.~P. Ivanov, {\em {Two-Higgs-doublet model from the group-theoretic
  perspective}}. \href{http://dx.doi.org/10.1016/j.physletb.2005.10.015}{Phys.
  Lett. {\bf B632} (2006)  360--365},
\href{http://arxiv.org/abs/hep-ph/0507132}{{\tt arXiv:hep-ph/0507132
  [hep-ph]}}.



\bibitem{Dev:2014yca}
P.~S. Bhupal~Dev and A.~Pilaftsis, {\em {Maximally Symmetric Two Higgs Doublet
  Model with Natural Standard Model Alignment}}.
  \href{http://dx.doi.org/10.1007/JHEP11(2015)147,
  10.1007/JHEP12(2014)024}{JHEP {\bf 12} (2014)  024},
  \href{http://arxiv.org/abs/1408.3405}{{\tt arXiv:1408.3405 [hep-ph]}}.
[Erratum: JHEP11,147(2015)].

\bibitem{Lane:2018ycs}
K.~Lane and W.~Shepherd, {\em {Natural Stabilization of the Higgs Boson's Mass
  and Alignment}}.
\href{http://arxiv.org/abs/1808.07927}{{\tt arXiv:1808.07927 [hep-ph]}}.
%

\bibitem{Bernon:2015qea}
J.~Bernon, J.~F. Gunion, H.~E. Haber, Y.~Jiang, and S.~Kraml, {\em
  {Scrutinizing the alignment limit in two-Higgs-doublet models: m$_h$=125
  GeV}}. \href{http://dx.doi.org/10.1103/PhysRevD.92.075004}{Phys. Rev. {\bf
  D92} (2015) no.~7, 075004},
\href{http://arxiv.org/abs/1507.00933}{{\tt arXiv:1507.00933 [hep-ph]}}.

\bibitem{Bernon:2015wef}
J.~Bernon, J.~F. Gunion, H.~E. Haber, Y.~Jiang, and S.~Kraml, {\em
  {Scrutinizing the alignment limit in two-Higgs-doublet models. II.
  m$_H$=125  GeV}}.
  \href{http://dx.doi.org/10.1103/PhysRevD.93.035027}{Phys. Rev. {\bf D93}
  (2016) no.~3, 035027},
\href{http://arxiv.org/abs/1511.03682}{{\tt arXiv:1511.03682 [hep-ph]}}.

\bibitem{Carena:2013ooa}
M.~Carena, I.~Low, N.~R. Shah, and C.~E.~M. Wagner, {\em {Impersonating the
  Standard Model Higgs Boson: Alignment without Decoupling}}.
  \href{http://dx.doi.org/10.1007/JHEP04(2014)015}{JHEP {\bf 04} (2014)  015},
\href{http://arxiv.org/abs/1310.2248}{{\tt arXiv:1310.2248 [hep-ph]}}.

\bibitem{Carena:2015moc}
M.~Carena, H.~E. Haber, I.~Low, N.~R. Shah, and C.~E.~M. Wagner, {\em
  {Alignment limit of the NMSSM Higgs sector}}.
  \href{http://dx.doi.org/10.1103/PhysRevD.93.035013}{Phys. Rev. {\bf D93}
  (2016) no.~3, 035013},
\href{http://arxiv.org/abs/1510.09137}{{\tt arXiv:1510.09137 [hep-ph]}}.

\bibitem{Haber:2017erd}
H.~E. Haber, S.~Heinemeyer, and T.~Stefaniak, {\em {The Impact of Two-Loop
  Effects on the Scenario of MSSM Higgs Alignment without Decoupling}}.
  \href{http://dx.doi.org/10.1140/epjc/s10052-017-5243-5}{Eur. Phys. J. {\bf
  C77} (2017) no.~11, 742},
\href{http://arxiv.org/abs/1708.04416}{{\tt arXiv:1708.04416 [hep-ph]}}.

\bibitem{Fayet:1975yi}
P.~Fayet, {\em {Fermi-Bose Hypersymmetry}}.
\href{http://dx.doi.org/10.1016/0550-3213(76)90458-2}{Nucl. Phys. {\bf B113}
  (1976)  135}.

\bibitem{delAguila:1984qs}
F.~del Aguila, M.~Dugan, B.~Grinstein, L.~J. Hall, G.~G. Ross, and P.~C. West,
  {\em {Low-energy Models With Two Supersymmetries}}.
\href{http://dx.doi.org/10.1016/0550-3213(85)90480-8}{Nucl. Phys. {\bf B250}
  (1985)  225--251}.

\bibitem{Antoniadis:2005em}
I.~Antoniadis, A.~Delgado, K.~Benakli, M.~Quiros, and M.~Tuckmantel, {\em
  {Splitting extended supersymmetry}}.
  \href{http://dx.doi.org/10.1016/j.physletb.2006.01.010}{Phys. Lett. {\bf
  B634} (2006)  302--306},
\href{http://arxiv.org/abs/hep-ph/0507192}{{\tt arXiv:hep-ph/0507192
  [hep-ph]}}.

\bibitem{Antoniadis:2006eb}
I.~Antoniadis, K.~Benakli, A.~Delgado, M.~Quiros, and M.~Tuckmantel, {\em
  {Split extended supersymmetry from intersecting branes}}.
  \href{http://dx.doi.org/10.1016/j.nuclphysb.2006.03.012}{Nucl. Phys. {\bf
  B744} (2006)  156--179},
\href{http://arxiv.org/abs/hep-th/0601003}{{\tt arXiv:hep-th/0601003
  [hep-th]}}.

\bibitem{Allanach:2006fy}
K.~Benakli and C.~Moura, {\em {Les Houches physics at TeV colliders 2005 beyond
  the standard model working group: Summary report}}.
\href{http://arxiv.org/abs/hep-ph/0602198}{{\tt arXiv:hep-ph/0602198
  [hep-ph]}}.

\bibitem{Fayet:1978qc}
P.~Fayet, {\em {Massive gluinos}}.
\href{http://dx.doi.org/10.1016/0370-2693(78)90474-4}{Phys. Lett. {\bf 78B}
  (1978)  417--420}.

\bibitem{Polchinski:1982an}
J.~Polchinski and L.~Susskind, {\em {Breaking of Supersymmetry at
  Intermediate-Energy}}.
\href{http://dx.doi.org/10.1103/PhysRevD.26.3661}{Phys. Rev. {\bf D26} (1982)
  3661}.

\bibitem{Hall:1990hq}
L.~J. Hall and L.~Randall, {\em {U(1)-R symmetric supersymmetry}}.
\href{http://dx.doi.org/10.1016/0550-3213(91)90444-3}{Nucl. Phys. {\bf B352}
  (1991)  289--308}.

\bibitem{Fox:2002bu}
P.~J. Fox, A.~E. Nelson, and N.~Weiner, {\em {Dirac gaugino masses and
  supersoft supersymmetry breaking}}.
  \href{http://dx.doi.org/10.1088/1126-6708/2002/08/035}{JHEP {\bf 08} (2002)
  035},
\href{http://arxiv.org/abs/hep-ph/0206096}{{\tt arXiv:hep-ph/0206096
  [hep-ph]}}.

\bibitem{Benakli:2008pg}
K.~Benakli and M.~D. Goodsell, {\em {Dirac Gauginos in General Gauge
  Mediation}}. \href{http://dx.doi.org/10.1016/j.nuclphysb.2009.03.002}{Nucl.
  Phys. {\bf B816} (2009)  185--203},
\href{http://arxiv.org/abs/0811.4409}{{\tt arXiv:0811.4409 [hep-ph]}}.

\bibitem{Benakli:2009mk}
K.~Benakli and M.~D. Goodsell, {\em {Dirac Gauginos and Kinetic Mixing}}.
  \href{http://dx.doi.org/10.1016/j.nuclphysb.2010.01.003}{Nucl. Phys. {\bf
  B830} (2010)  315--329},
\href{http://arxiv.org/abs/0909.0017}{{\tt arXiv:0909.0017 [hep-ph]}}.

\bibitem{Amigo:2008rc}
S.~D.~L. Amigo, A.~E. Blechman, P.~J. Fox, and E.~Poppitz, {\em {R-symmetric
  gauge mediation}}.
  \href{http://dx.doi.org/10.1088/1126-6708/2009/01/018}{JHEP {\bf 01} (2009)
  018},
\href{http://arxiv.org/abs/0809.1112}{{\tt arXiv:0809.1112 [hep-ph]}}.

\bibitem{Benakli:2010gi}
K.~Benakli and M.~D. Goodsell, {\em {Dirac Gauginos, Gauge Mediation and
  Unification}}. \href{http://dx.doi.org/10.1016/j.nuclphysb.2010.06.018}{Nucl.
  Phys. {\bf B840} (2010)  1--28},
\href{http://arxiv.org/abs/1003.4957}{{\tt arXiv:1003.4957 [hep-ph]}}.

\bibitem{Choi:2010gc}
S.~Y. Choi, D.~Choudhury, A.~Freitas, J.~Kalinowski, J.~M. Kim, and P.~M.
  Zerwas, {\em {Dirac Neutralinos and Electroweak Scalar Bosons of N=1/N=2
  Hybrid Supersymmetry at Colliders}}.
  \href{http://dx.doi.org/10.1007/JHEP08(2010)025}{JHEP {\bf 08} (2010)  025},
\href{http://arxiv.org/abs/1005.0818}{{\tt arXiv:1005.0818 [hep-ph]}}.

\bibitem{Benakli:2011vb}
K.~Benakli, {\em {Dirac Gauginos: A User Manual}}.
  \href{http://dx.doi.org/10.1002/prop.201100071}{Fortsch. Phys. {\bf 59}
  (2011)  1079--1082},
\href{http://arxiv.org/abs/1106.1649}{{\tt arXiv:1106.1649 [hep-ph]}}.

\bibitem{Benakli:2011kz}
K.~Benakli, M.~D. Goodsell, and A.-K. Maier, {\em {Generating mu and Bmu in
  models with Dirac Gauginos}}.
  \href{http://dx.doi.org/10.1016/j.nuclphysb.2011.06.001}{Nucl. Phys. {\bf
  B851} (2011)  445--461},
\href{http://arxiv.org/abs/1104.2695}{{\tt arXiv:1104.2695 [hep-ph]}}.

\bibitem{Itoyama:2011zi}
H.~Itoyama and N.~Maru, {\em {D-term Dynamical Supersymmetry Breaking
  Generating Split N=2 Gaugino Masses of Mixed Majorana-Dirac Type}}.
  \href{http://dx.doi.org/10.1142/S0217751X1250159X}{Int. J. Mod. Phys. {\bf
  A27} (2012)  1250159},
\href{http://arxiv.org/abs/1109.2276}{{\tt arXiv:1109.2276 [hep-ph]}}.

\bibitem{Benakli:2012cy}
K.~Benakli, M.~D. Goodsell, and F.~Staub, {\em {Dirac Gauginos and the 125 GeV
  Higgs}}. \href{http://dx.doi.org/10.1007/JHEP06(2013)073}{JHEP {\bf 06}
  (2013)  073},
\href{http://arxiv.org/abs/1211.0552}{{\tt arXiv:1211.0552 [hep-ph]}}.

\bibitem{Benakli:2014cia}
K.~Benakli, M.~Goodsell, F.~Staub, and W.~Porod, {\em {Constrained minimal
  Dirac gaugino supersymmetric standard model}}.
  \href{http://dx.doi.org/10.1103/PhysRevD.90.045017}{Phys. Rev. {\bf D90}
  (2014) no.~4, 045017},
\href{http://arxiv.org/abs/1403.5122}{{\tt arXiv:1403.5122 [hep-ph]}}.

\bibitem{Martin:2015eca}
S.~P. Martin, {\em {Nonstandard supersymmetry breaking and Dirac gaugino masses
  without supersoftness}}.
  \href{http://dx.doi.org/10.1103/PhysRevD.92.035004}{Phys. Rev. {\bf D92}
  (2015) no.~3, 035004},
\href{http://arxiv.org/abs/1506.02105}{{\tt arXiv:1506.02105 [hep-ph]}}.

\bibitem{Braathen:2016mmb}
J.~Braathen, M.~D. Goodsell, and P.~Slavich, {\em {Leading two-loop corrections
  to the Higgs boson masses in SUSY models with Dirac gauginos}}.
  \href{http://dx.doi.org/10.1007/JHEP09(2016)045}{JHEP {\bf 09} (2016)  045},
\href{http://arxiv.org/abs/1606.09213}{{\tt arXiv:1606.09213 [hep-ph]}}.

\bibitem{Unwin:2012fj}
J.~Unwin, {\em {R-symmetric High Scale Supersymmetry}}.
  \href{http://dx.doi.org/10.1103/PhysRevD.86.095002}{Phys. Rev. {\bf D86}
  (2012)  095002},
\href{http://arxiv.org/abs/1210.4936}{{\tt arXiv:1210.4936 [hep-ph]}}.

\bibitem{Chakraborty:2018izc}
S.~Chakraborty, A.~Martin, and T.~S. Roy, {\em {Charting generalized supersoft
  supersymmetry}}. \href{http://dx.doi.org/10.1007/JHEP05(2018)176}{JHEP {\bf
  05} (2018)  176},
\href{http://arxiv.org/abs/1802.03411}{{\tt arXiv:1802.03411 [hep-ph]}}.

\bibitem{Csaki:2013fla}
C.~Csaki, J.~Goodman, R.~Pavesi, and Y.~Shirman, {\em {The $m_D-b_M$ problem of
  Dirac gauginos and its solutions}}.
  \href{http://dx.doi.org/10.1103/PhysRevD.89.055005}{Phys. Rev. {\bf D89}
  (2014) no.~5, 055005},
\href{http://arxiv.org/abs/1310.4504}{{\tt arXiv:1310.4504 [hep-ph]}}.

\bibitem{Benakli:2016ybe}
K.~Benakli, L.~Darm\'e, M.~D. Goodsell, and J.~Harz, {\em {The Di-Photon Excess
  in a Perturbative SUSY Model}}.
  \href{http://dx.doi.org/10.1016/j.nuclphysb.2016.07.027}{Nucl. Phys. {\bf
  B911} (2016)  127--162},
\href{http://arxiv.org/abs/1605.05313}{{\tt arXiv:1605.05313 [hep-ph]}}.


%

\bibitem{Haber:1993an}
H.~E. Haber and R.~Hempfling, {\em {The Renormalization group improved Higgs
  sector of the minimal supersymmetric model}}.
  \href{http://dx.doi.org/10.1103/PhysRevD.48.4280}{Phys. Rev. {\bf D48} (1993)
   4280--4309},
\href{http://arxiv.org/abs/hep-ph/9307201}{{\tt arXiv:hep-ph/9307201
  [hep-ph]}}.



\end{thebibliography}
\end{document}